\documentstyle[prc,aps,epsf]{revtex}
\begin{document}
\title {Short-range correlations in
low-lying nuclear excited states} 

\author
{S.R. Mokhtar$^{1,3}$, G. Co'$^{1,2}$ and A. M. Lallena$^4$ }
\address{
$^{1)}$Dipartimento di Fisica, Universit\`a di Lecce,
        I-73100 Lecce, Italy \\ 
$^{2)}$ Istituto Nazionale di Fisica Nucleare sez. di Lecce,
        I-73100 Lecce, Italy \\
$^{3)}$ Department of Physics, University of Assiut,
        Assiut, Egypt \\ 
$^{4)}$Departamento de F\'{\i}sica
Moderna, Universidad de Granada, 
E-18071 Granada, Spain}
\maketitle
\begin{abstract}
The electromagnetic transitions to various low-lying excited states of
$^{16}$O, $^{48}$Ca and $^{208}$Pb are calculated within a model
which considers the short-range correlations. 
In general the effects of the correlations are small
and do not explain the required quenching to describe the data.
\end{abstract} 
%
%
\vskip .3cm
PACS number(s): 21.60.-n, 25.30.Dh

\vskip 1.cm

The description of the electron scattering
form factors of low-lying excited states is still an unsolved problem.
The disagreement between theory and experiment is particularly
annoying for the high angular momentum stretched states
which are composed by few particle-hole (p-h) configurations. Because
of this relatively simple structure one expects that the
independent particle model (IPM) should be able to reproduce the
data. This is not far to be true 
for what concerns the shape of the form factors, but
the theoretical results usually overestimate the data. 
The disagreement between theory and experiment is commonly summarized by 
a single number: the quenching factor Q, necessary to reproduce the
data. 

There have been various attempts to identify the sources of this
disagreement, but the situation has not yet been clarified. 
The quenching produced by first order core polarization mechanism has
been proposed in Ref. \cite{ham80} but other studies \cite{suz83}
found this effect to be small.
The quenchings of the 12$^-$ and 14$^-$ form factors in $^{208}$Pb
have been explained by a Random Phase Approximation (RPA) plus
particle vibration coupling model in Ref. \cite{kre80} 
but these results have not been confirmed by a
self consistent first- and second- RPA calculation \cite{dec81}.

In Ref. \cite{pan84} a mechanism related to the presence of
short-range correlations has been proposed. These correlations modify
the occupation probability of the single particle levels, reducing the
occupation of the hole states and producing a finite probability of
occupying the particle states even when the nucleus is in its ground
state. This idea is supported by the fact that elastic electron
scattering data \cite{cav82} and (e,e'p) data \cite{qui88} in
$^{208}$Pb can be explained assuming partial occupation probability of
the single particle levels closed the Fermi surface \cite{pap86}.

In this work we have studied the electron excitation of some non
collective low-lying states of doubly closed shell nuclei with a model
that takes into account the short-range correlations.  The starting
point of our model is the CBF ansatz for the description of the
nuclear ground state:
\begin{equation}
|\Psi_0 \rangle \,= \,G \,|\Phi_0 \rangle \, ,
\label{psi0def}
\end{equation}
where $G$ is a correlation function and $|\Phi \rangle$ is a Slater
determinant of single particle wave functions. 
The many-body responses induced by an operator $O({\bf q})$ can be
written as:
\begin{eqnarray}
\nonumber
S({\bf q} ,\omega) \, &=& \, - \frac{Im}{\pi}
\frac {
\langle \Psi_0 |\,O^{+}({\bf q})\,
(H - E_0 - \omega + i \eta)^{-1}\,O ({\bf q}) \,
| \Psi_0 \rangle  \, }
{ \langle \Psi_0 | \Psi_0 \rangle } \\
 &=&\,
- \frac{Im}{\pi} \sum_n \, \xi_n^+({\bf q})\, 
(E_n - E_0 - \omega + i \eta)^{-1} \, \xi_n({\bf q})
\, ,
\label{resp}
\end{eqnarray}
where:
\begin{equation}
\xi_n({\bf q})\, =\, 
\frac { \langle \Psi_n | \,O({\bf q})\, |\Psi_0 \rangle }
 {\langle \Psi_n |\Psi_n\rangle ^{\frac{1}{2}}\,
\langle \Psi_0 |\Psi_0\rangle ^{\frac{1}{2}} }\, .
\label{xi1}
\end{equation}
In the above equations $H$ indicates the hamiltonian, ${\bf q}$ and
$\omega$ the momentum and energy transfer respectively and $E$ the
energies of the nuclear system.

In our model the excited states are defined in analogy with
Eq. (\ref{psi0def}):  $|\Psi_n \rangle = G |\Phi_n \rangle $. 
The correlation $G$ is the same used to describe the ground state.
With this ansatz we rewrite Eq. (\ref{xi1}) as:
\begin{equation}
\xi_n({\bf q}) \,= \,
\frac {\langle \Phi_n |\,G^+ \,O({\bf q}) \,G|\Phi_0 \rangle
}{\langle \Phi_0|\,G^+\,G\,|\Phi_0\rangle } \left[ \frac {\langle \Phi_0
|\,G^+\,G\,|\Phi_0\rangle }{\langle \Phi_n|\,G^+\,G\,|\Phi_n\rangle }
\right]^\frac{1}{2} \, .
\label{xi2}
\end{equation}

In principle
the correlation function $G$ has a complicated operatorial structure,
analogous to that of the nuclear hamiltonian. In the present work we
have considered a purely scalar correlation function which is 
therefore commuting with the excitation operator $O({\bf q})$.
The functional dependence of the correlation function is 
the Jastrow ansatz \cite{jas55}:
\begin{equation}
G(1,2...A)\,=\,  \prod_{i<j}\, f(r_{ij})  \, ,
\label{cordef}
\end{equation}
where $f$ is a two-body correlation function and $r_{ij}$ is the
distance between two nucleons.  We introduce a function
$h_{ij}=f^2_{ij}-1$ and we perform a cluster expansion \cite{fan87} 
retaining only
those terms where the function $h_{ij}$ appears only once.
We have tested the model by comparing
its results with those of the full calculations. A first
comparison has been done for the nuclear matter quasi-elastic
charge responses \cite{ama98} and has shown the
validity of our approximation. The model has also been used
to evaluate the ground state
charge density distributions \cite{co95,ari97} of various doubly closed
shell nuclei. Even in this case the
agreement with the results of the full calculations is very good.   

In the present work we apply our model to describe the excitation of
low-lying states induced by electron scattering.
A detailed description of the model can be found 
in Ref. \cite{co00} where the explicit expressions of the transition
matrix elements are given. 
We should remark however that while in the calculations of
Ref. \cite{co00} the convection current was considered only at the
mean-field level, in the present
calculations, we have also evaluated all the correlated terms related to
this current.

Our model requires two inputs, the set of single particle wave functions
and the correlation function. 
They are not independent since, for a given hamiltonian, 
they are fixed through the variational principle 
by minimizing the energy expectation value. 
The single particle wave functions and the correlations
we have used have been 
taken from Ref. \cite{ari96} where energy minima of several
doubly magic nuclei for the semi-realistic S3
interaction of Afnan and Tang have been found.

With this input we have calculated the electron excitations of
various low-lying states in $^{16}$O, $^{48}$Ca and
$^{208}$Pb nuclei and we have compared our results with the
experimental data of Refs. \cite{hyd84}-\cite{con92}. 

For magnetic excitations the comparison is done with 
the transverse form factor
which can be unambiguously obtained for each cross
section value. 
In the electric excitations, both longitudinal and transverse
form factors contribute. Their extraction from the cross section data
requires at least two measurements done at same value of the momentum
transfer ${\bf q}$. Because of the difficulties related to
this procedure \cite{co87}, for the electric states
we preferred to compare our calculations directly to the cross
sections. 

Since our model cannot describe collective effects, 
we have considered only those states characterized by one or at most
two  p-h excitations. 
We have estimated the degree of collectivity of the various excited
states by making discrete RPA calculations with both
a density dependent
Landau-Migdal interaction \cite{rin78} and the J\"ulich-Stony Brook
interaction \cite{spe80}. We have selected those states having at
least one p-h transition with $X$ amplitude value
larger than 0.9 in both calculations.

From this analysis we found three states 
dominated by two p-h pairs. 
For these cases we have supposed that the wave
function of the excited uncorrelated state could be described as a
linear combination of the two p-h pairs:
\begin{equation}
\label{rpax}
|\Phi_n \rangle \,= \, X_{ph} \,|\Phi_{ph} \rangle \, 
                  + \, X_{p'h'} \,|\Phi_{p'h'} \rangle \, , 
\end{equation}
where the $X_{ph}$ amplitudes have been taken from the RPA solution and
the other one have been fixed such that $ X^2_{ph}+ X^2_{p'h'}=1$.

The results of our calculations are summarized in
Table \ref{tab:quenrt}. The quenching factors $Q$ for the uncorrelated
(IPM) and correlated calculations and the $\chi^2$ per datum are
compared. The values of the $\chi^2$ have been evaluated after the
application of the quenching factors to the original results.

The short-range correlations do
not substantially change the IPM results. In some
cases, the correlations reduce the cross section values, 
therefore the quenching factors increase.
There are, however,
various situations where the correlations effects go in the opposite
direction.  

We show in Figs. 1 and 2 the results obtained  
for the 4$^-$ state in $^{16}$O and for the 12$^+$ in $^{208}$Pb 
to make more explicit some of the features of the results. 
The linear scale used in some panel of the
two figures allows for a better identification of the correlations
effects consisting in a more or less pronounced modification of the
maximum of the distribution. The
results are also plotted in a logarithmic scale since they
are commonly presented in this way. 
In the  4$^-$ case the correlations
enhance the value of the maximum. This implies a smaller  $Q$
for the  correlated calculation, as it is shown in the table.
Even after the $Q$ factor has been applied
the agreement with the data is rather
poor, as the high values of the $\chi^2$ indicate.

In Fig. 2 the 12$^+$ cross sections calculated for two
different values of the scattering angle are presented and compared
with the data of Ref. \cite{lic79}.  In this case the correlation
lowers the maximum of the distribution.  The 12$^+$ excitation has
been calculated as a pure neutron 1i$_{11/2}$~1i$_{13/2}^{-1}$
transition. The difference in the lowering produced by the correlation
in panels $a$ and $b$ of the figure, 
i.e. the ratio between full and dashed lines, 
is within the 1\%. This small
difference is produced by the presence of the longitudinal
response generated by the electric neutron form factor 
of Ref. \cite{hoe76} which we use in our calculations. 

One can notice that the quenching factors required to reproduce the
data are different for the two different scattering angles.
However these two values are statistically compatible
once the experimental uncertainties are considered.

From what we have presented it appears evident that the
correlation effects are very small and cannot
be considered relevant for the description of the experimental data.
The size of these effects
is in agreement with the results of
microscopic nuclear matter calculations done within the Correlated
Basis Function theory and using the Fermi Hypernetted Chain
resummation techniques \cite{ben90} which find
occupation numbers very close to those of the IPM. 

It is possible that some of the correlation components we have
neglected, especially the tensor terms,
could produce noticeable effects, 
as indicated in Ref. \cite{fab97}.
On the other hand, microscopic
calculations in both infinite systems \cite{wir88}
and finite nuclei \cite{fab00}, show that the scalar term is by far the
largest one in the correlation.
Furthermore, a study of the ground state charge and momentum
distributions of doubly closed shell nuclei \cite{ari97} indicate the
small influence of these state dependent terms.

Our experience in RPA calculations \cite{lal88}
has shown that the presence of
small amplitude p-h pairs can heavily modify the size, and sometime
even the shape, of the form factor. 
For this reason we think that the origin of the quenching
factor should be searched by looking with more detail
at the coupling of the single
particle excitations with the collective modes of the nucleus.

This work has been partially supported by the CICYT-INFN agreement and
by the DGES (PB98-1367) and the Junta de Andaluc\'{\i}a (FQM225).

%
%
%
%
\begin{table}
\caption{ Excited states calculated and their quenching factors. We
have indicated the excitation energy $E$, the p-h pairs considered, the
$X$ amplitudes, Eq. (\protect\ref{rpax}), the quenching factors $Q$ and the
$\chi^2$ for both calculations, and the reference where the
experimental points have been taken. For the electric states we have
considered the quenching on the full cross section, and we have
indicated the scattering angle were the cross section has been measured.}
\label{tab:quenrt}
\begin{center}
\begin{tabular}{ccrccrrrrc}
   & J$^{\pi}$ & $E$ (MeV) & p-h pairs & $X_{ph}$ &
\multicolumn{2}{c}{IPM} & \multicolumn{2}{c}{Correlated} & Ref. \\
         &  & & & & $Q$ & $\chi^2$ & $Q$ & $\chi^2$ & \\
\hline
\rule{0cm}{.3cm}$^{16}$O & 4$^-$ & 18.98 & $\pi$ 1d$_{5/2}$ 1p$_{3/2}^{-1}$ & +0.7058
&&&&&\\[-.2cm]
&&&&& 0.67 & 26.35 & 0.66 & 35.06 & \cite{hyd84}  \\[-.2cm]
         &  &     &$\nu$ 1d$_{5/2}$ 1p$_{3/2}^{-1}$ & -0.7084  & & & & & \\
\hline
\rule{0cm}{.3cm}$^{48}$Ca& 4$^-$ & 6.105  &$\pi$ 1f$_{7/2}$ 2s$_{1/2}^{-1}$ & -0.7520 
&&&&&\\[-.2cm]
&&&&                 & 0.61 & 16.95 & 0.61 & 16.98 & \cite{wis85}\\[-.2cm]
         &  &     &$\pi$ 1f$_{7/2}$ 1d$_{3/2}^{-1}$ & -0.6591 &  & & & & \\
         & 6$^-$ & 8.557  &$\pi$ 1f$_{7/2}$ 1d$_{5/2}^{-1}$ & 1.0 
                 & 0.06 & 6.99 & 0.06 & 8.67 & \cite{wis85}  \\
         & 8$^-$ & 9.276  &$\nu$ 1g$_{9/2}$ 1f$_{7/2}^{-1}$ & 1.0 
                 & 0.26 & 6.39 & 0.25 & 7.94 & \cite{wis85}  \\
         & 3$^+$ & 4.608 &$\nu$ 2p$_{3/2}$ 1f$_{7/2}^{-1}$ & 1.0 
                 & 0.32 & 1.93 & 0.34 & 2.26 & \cite{wis85}  \\
         & 5$^+$ & 5.147 &$\nu$ 2p$_{3/2}$ 1f$_{7/2}^{-1}$ & 1.0 
                 & 0.42 & 20.77 & 0.43 & 19.83 & \cite{wis85}  \\
$\theta=160^0$ 
        & 5$^-$ & 8.804 &$\nu$ 1g$_{9/2}$ 1f$_{7/2}^{-1}$ & -0.8242 
&&&&&\\[-.2cm]
&&&&                 & 0.34 & 7.11 & 0.37 & 8.01 & \cite{wis85}  \\[-.2cm]
         &  &     &$\pi$ 1f$_{7/2}$ 1d$_{3/2}^{-1}$ & +0.5663 &  & & & & \\
\hline
\rule{0cm}{.3cm}$^{208}$Pb& 9$^+$ & 5.01  &$\nu$ 2g$_{9/2}$ 1i$_{13/2}^{-1}$ & 1.0 
                 & 0.38 & 5.92 & 0.37 & 7.17 & \cite{con92}  \\
          & 9$^+$ & 5.26  &$\pi$ 1h$_{9/2}$ 1h$_{11/2}^{-1}$ & 1.0 
                 & 0.59 & 2.09 & 0.59 & 2.10 & \cite{con92}  \\
          &10$^-$ & 6.283  &$\nu$ 1j$_{15/2}$ 1i$_{13/2}^{-1}$ & 1.0 
                 & 0.34 & 9.12 & 0.34 & 9.29 & \cite{con92}  \\
          &10$^-$ & 6.884  &$\pi$ 1i$_{13/2}$ 1h$_{11/2}^{-1}$ & 1.0 
                 & 0.33 & 6.05 & 0.33 & 6.05 & \cite{con92}  \\
          &12$^-$ & 6.437  &$\nu$ 1j$_{15/2}$ 1i$_{13/2}^{-1}$ & 1.0 
                 & 0.70 & 3.35 & 0.68 & 3.57 & \cite{con92}  \\
          &12$^-$ & 7.064  &$\pi$ 1i$_{13/2}$ 1h$_{11/2}^{-1}$ & 1.0 
                 & 0.28 & 7.64 & 0.27 & 8.85 & \cite{con92}  \\
          &14$^-$ & 6.745  &$\nu$ 1j$_{15/2}$ 1i$_{13/2}^{-1}$ & 1.0 
                 & 0.39 & 5.53 & 0.39 & 5.53 & \cite{con92}  \\
$\theta=90^0$ 
        & 10$^+$ & 5.920 &$\nu$ 1i$_{11/2}$ 1i$_{13/2}^{-1}$ & 1.0 
                 & 0.63 & 18.90 & 0.69 & 20.63 & \cite{lic79}  \\
$\theta=160^0$ 
        & 10$^+$   & 5.920 &$\nu$ 1i$_{11/2}$ 1i$_{13/2}^{-1}$ & 1.0  
                 & 0.88 & 28.91 & 0.95 & 32.15 & \cite{lic79}  \\
$\theta=90^0$ 
        & 12$^+$ & 6.100 &$\nu$ 1i$_{11/2}$ 1i$_{13/2}^{-1}$ & 1.0 
                 & 0.52 & 7.55 & 0.57 & 8.76 & \cite{lic79}  \\
$\theta=160^0$ 
        &  12$^+$ & 6.100 &$\nu$ 1i$_{11/2}$ 1i$_{13/2}^{-1}$  &  1.0  
                 & 0.39 & 12.70 & 0.42 & 14.84 & \cite{lic79}  \\
\end{tabular}
\end{center}
\end{table}
%

%
%
%
%
\newpage
\begin{figure}
\vskip 1 cm
\noindent
\begin{center}
\vspace*{3.cm}
\leavevmode
\epsfysize = 350pt
\epsfbox[70 200 500 650]{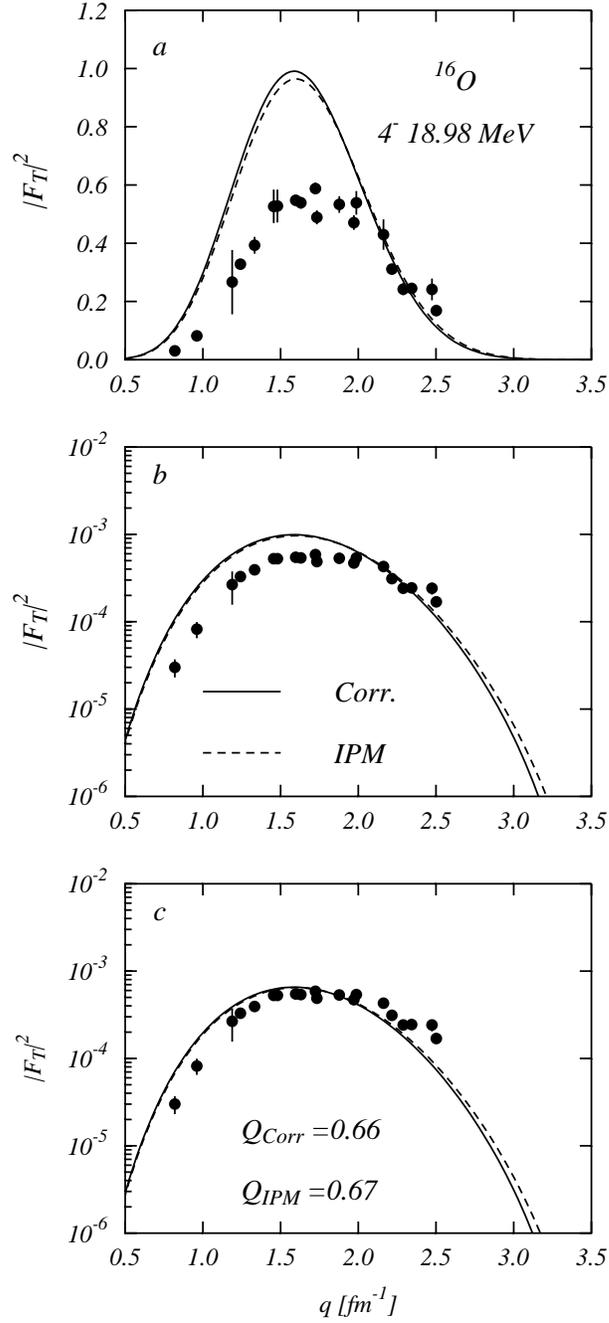}
\end{center}
\vspace{4.0cm}
\caption{Transverse form factor for the 4$^-$ in $^{16}$O
state at 18.98
MeV. In the panel $a$ the results of the IPM (dashed line) and of our
correlated model (full line) are compared with the data of
Ref. \protect\cite{hyd84}. The same figure is redrawn in panel $b$ in log
scale since these results are commonly presented in this way. 
The panel $c$ show the
comparison between theoretical results and experimental data, after
the quenching factors have been applied.}
\label{fig:o16}
\end{figure}
%
%
\newpage
\begin{figure}
\begin{center}
\vspace*{3.cm}
\leavevmode
\epsfysize = 350pt
\epsfbox[70 200 500 650]{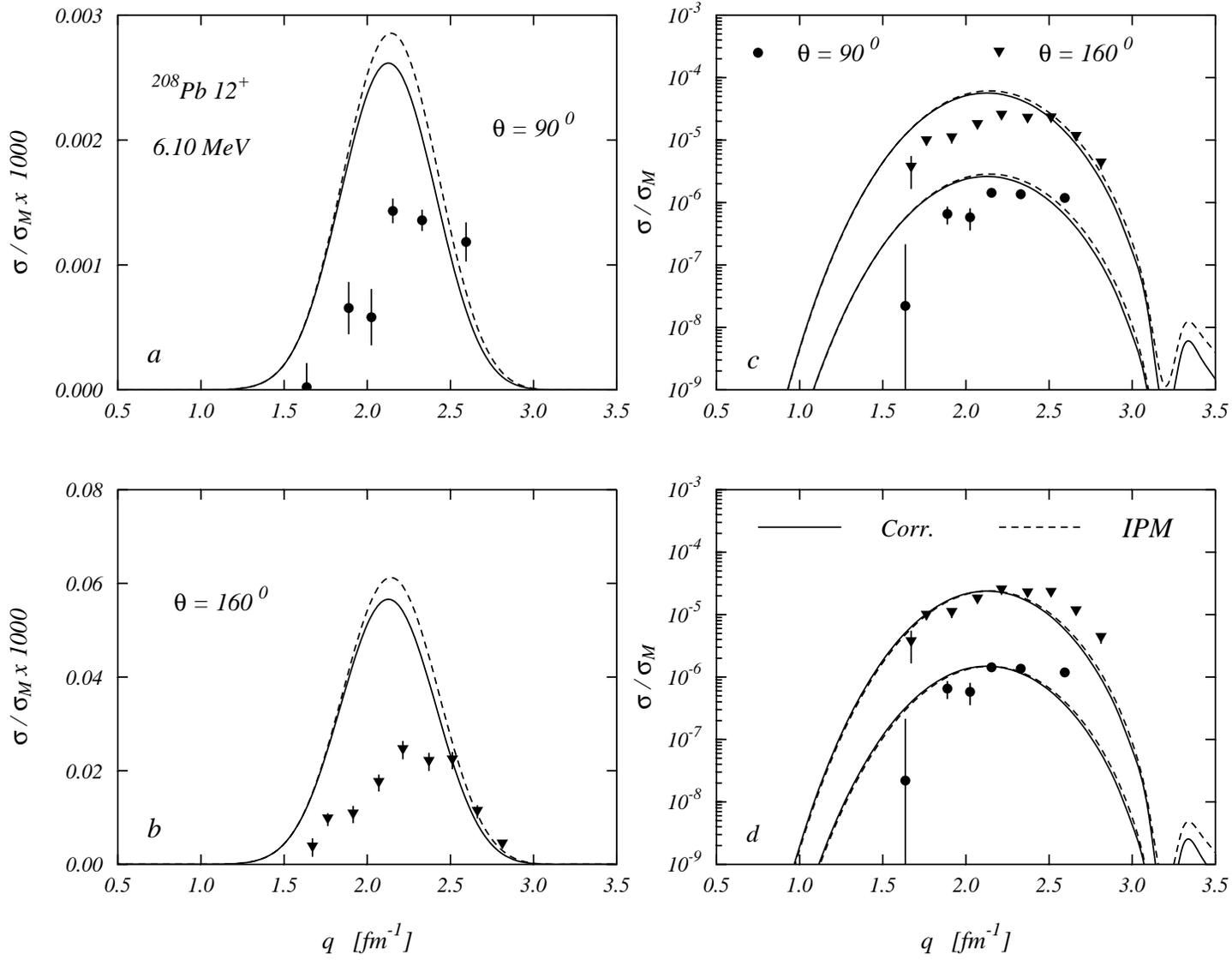}
\end{center}
\vspace{4.5cm}
\caption{Cross sections divided by the Mott cross section
$\sigma_M$ for the 12$^+$ state in $^{208}$Pb at 6.1 MeV measured at
two different scattering angles. In the panel $a$ and $b$ the
comparison of the IPM (dashed line) and correlated (full line) results
with the data of Ref. \protect\cite{lic79} is done in linear
scale. The same comparison drawn in log scale is presented in the
panel c. In the panel $d$ the comparison is done after the application
of the quenching factors.}
\label{fig:pb208}
\end{figure}
\end{document}